\documentclass[aps,prl,showpacs,twocolumn]{revtex4}

\usepackage{amsmath}
\usepackage{graphicx}
\usepackage{dcolumn}
\usepackage{bm}

\begin{document}

{\bf Reply to Comment by Calandra \textit{et al} on \lq\lq{}Electronic Structure of Superconducting KC$_8$ and Nonsuperconducting LiC$_6$ Graphite Intercalation Compounds: Evidence for a Graphene-Sheet-Driven Superconducting State\rq\rq{}}

In their comment Calandra \textit{et al} \cite{Calandra}, assert two points: (1) the estimate of charge transfer from Li to graphene layers in LiC$_6$ in our letter \cite{Pan2011c} is incorrect because of the three dimensional (3D) character of the electronic structure in bulk LiC$_6$; (2) our main claim that the superconductivity in graphite intercalation compounds (GICs) is graphene-sheet-driven is therefore invalid.
 
First, we point out that our claim on graphene driven superconductivity in GICs is based on the experimental results from a whole series of different materials (graphite, KC$_24$, LiC$_6$, KC$_8$ and CaC$_6$) and that it is valid regardless of the charge transfer estimate. In these different GICs, we observe a strong electron phonon coupling (EPC) between the \textbf{graphene derived electrons and graphene derived phonons} \cite{Valla2009,Pan2011c}. When put in the McMillan's formula, the measured coupling constants give the superconducting transition temperatures, T$_c$, that are very close to the measured ones in LiC$_6$, KC$_8$ and CaC$_6$, demonstrating that the graphene sheets are indeed crucial for superconductivity in GICs. The side observation that the filling of the $\pi^\star$ states follows the same trend is in accord with a simple picture where the EPC strengthens as the phase space for the scattering grows with the size of the Fermi surface. However, this observation is not essential for the main conclusion of our letter. 
Second, we note that the validity of the calculations and the estimate for the charge transfer in Calandra \textit{et al} is heavily based on comparison with the data from another material, lithium intercalated graphene bi-layer \cite{Sugawara2011}, irrelevant for the studies of bulk GICs. 

The third and the most important point is that the calculations for LiC$_6$ show essentially a 3D electronic structure, virtually unchanged from the early work by Holzwarth \textit{et al} \cite{Holzwarth1977}, whereas our photoemission experiments show no out-of -plane dispersion. Fig. 1 shows the $\pi^\star$-derived Fermi surface (FS) of LiC$_6$ recorded at different photon energies from samples with larger crystallites and a higher degree of intercalant order than those from Pan \textit{et al} \cite{Pan2011c}. The three contours, originating from the A$\alpha$A$\beta$A$\gamma$ stacking in LiC$_6$ below 220 K \cite{Kambe1979}, are now clearly visible, indicating perfect stacking. The relative intensity of these three contours varies, but their areas do not change with $k_z$. As the FS contours are sharper than in ref. \cite{Pan2011c}, the charge transfer could be more precisely determined: the FS area is somewhat larger than in ref. \cite{Pan2011c}, corresponding to the charge transfer of 0.052 e$^-$ per graphene unit cell (GUC), still significantly smaller than in KC$_8$ (0.11 e$^-$/GUC). The momentum averaged EPC is the same, within the error bars, to the value reported in Pan \textit{et al} \cite{Pan2011c}.
The lack of $k_z$ dispersion in the experiments clearly demonstrates inability of DFT calculations to correctly describe LiC$_6$. A possible reason might be the wrong crystal structure - A$\alpha$, instead of the correct A$\alpha$A$\beta$A$\gamma$ stacking has always been used as a starting point in these calculations.
However, for the problem of superconductivity in GICs, the more consequential issue is inability of the existing DFT calculations to account for the enhancement of the EPC on the $\pi^\star$-derived Fermi surface in GICs with doping, observed in many experiments, including ours \cite{Valla2009,Pan2011c}.

\begin{figure}[tb]
\begin{center}
\includegraphics[width=.85\columnwidth,angle=0]{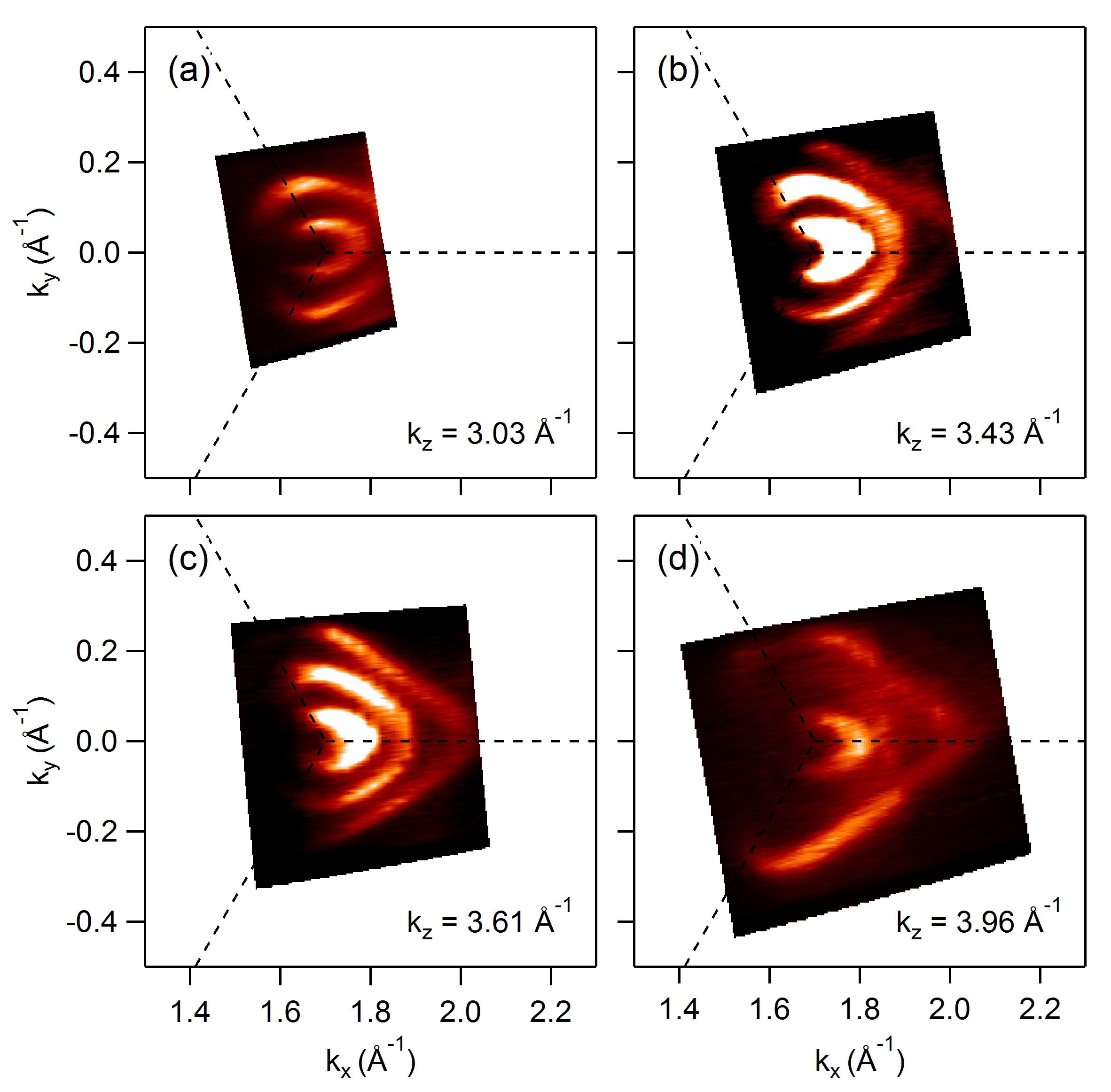}
\end{center}
\caption{Fermi surface of LiC$_6$ measured at $T=15$ K at four different photon energies: a) 40 eV, b) 50 eV, c) 55 eV and d) 65 eV. Corresponding $k_z$ values are indicated.}
\end{figure}

Z.-H. Pan$^{1}$, A. V. Fedorov$^{2}$, C. A. Howard$^{3}$, M. Ellerby$^{3}$ and T. Valla$^1$
\\
$^{1}$Condensed Matter Physics and Materials Science Department, Brookhaven National Lab, Upton, NY 11973\\
$^{2}$Advanced Light Source, Lawrence Berkeley National Laboratory, Berkeley, CA 94720\\
$^{3}$London Centre for Nanotechnology and Department of Physics and Astronomy, University College London, London WC1E 6BT, United Kingdom


\begin{thebibliography}{6}
\expandafter\ifx\csname natexlab\endcsname\relax\def\natexlab#1{#1}\fi
\expandafter\ifx\csname bibnamefont\endcsname\relax
  \def\bibnamefont#1{#1}\fi
\expandafter\ifx\csname bibfnamefont\endcsname\relax
  \def\bibfnamefont#1{#1}\fi
\expandafter\ifx\csname citenamefont\endcsname\relax
  \def\citenamefont#1{#1}\fi
\expandafter\ifx\csname url\endcsname\relax
  \def\url#1{\texttt{#1}}\fi
\expandafter\ifx\csname urlprefix\endcsname\relax\def\urlprefix{URL }\fi
\providecommand{\bibinfo}[2]{#2}
\providecommand{\eprint}[2][]{\url{#2}}

\bibitem[{\citenamefont{Calandra et~al.}()\citenamefont{Calandra, Attaccalite,
  Profeta, and Mauri}}]{Calandra}
\bibinfo{author}{\bibfnamefont{M.}~\bibnamefont{Calandra}},
  \bibinfo{author}{\bibfnamefont{C.}~\bibnamefont{Attaccalite}},
  \bibinfo{author}{\bibfnamefont{G.}~\bibnamefont{Profeta}}, \bibnamefont{and}
  \bibinfo{author}{\bibfnamefont{F.}~\bibnamefont{Mauri}},
  \bibinfo{journal}{Phys. Rev. Lett.}  (unpublished).

\bibitem[{\citenamefont{Pan et~al.}(2011)\citenamefont{Pan, Camacho, Upton,
  Fedorov, Howard, Ellerby, and Valla}}]{Pan2011c}
\bibinfo{author}{\bibfnamefont{Z.-H.} \bibnamefont{Pan}},
  \bibinfo{author}{\bibfnamefont{J.}~\bibnamefont{Camacho}},
  \bibinfo{author}{\bibfnamefont{M.}~\bibnamefont{Upton}},
  \bibinfo{author}{\bibfnamefont{A.}~\bibnamefont{Fedorov}},
  \bibinfo{author}{\bibfnamefont{C.}~\bibnamefont{Howard}},
  \bibinfo{author}{\bibfnamefont{M.}~\bibnamefont{Ellerby}}, \bibnamefont{and}
  \bibinfo{author}{\bibfnamefont{T.}~\bibnamefont{Valla}},
  \bibinfo{journal}{Phys. Rev. Lett.} \textbf{\bibinfo{volume}{106}}
  (\bibinfo{year}{2011}).

\bibitem[{\citenamefont{Valla et~al.}(2009)\citenamefont{Valla, Camacho, Pan,
  Fedorov, Walters, Howard, and Ellerby}}]{Valla2009}
\bibinfo{author}{\bibfnamefont{T.}~\bibnamefont{Valla}},
  \bibinfo{author}{\bibfnamefont{J.}~\bibnamefont{Camacho}},
  \bibinfo{author}{\bibfnamefont{Z.-H.} \bibnamefont{Pan}},
  \bibinfo{author}{\bibfnamefont{A.}~\bibnamefont{Fedorov}},
  \bibinfo{author}{\bibfnamefont{A.}~\bibnamefont{Walters}},
  \bibinfo{author}{\bibfnamefont{C.}~\bibnamefont{Howard}}, \bibnamefont{and}
  \bibinfo{author}{\bibfnamefont{M.}~\bibnamefont{Ellerby}},
  \bibinfo{journal}{Phys. Rev. Lett.} \textbf{\bibinfo{volume}{102}}
  (\bibinfo{year}{2009}).

\bibitem[{\citenamefont{Sugawara et~al.}(2011)\citenamefont{Sugawara, Kanetani,
  Sato, and Takahashi}}]{Sugawara2011}
\bibinfo{author}{\bibfnamefont{K.}~\bibnamefont{Sugawara}},
  \bibinfo{author}{\bibfnamefont{K.}~\bibnamefont{Kanetani}},
  \bibinfo{author}{\bibfnamefont{T.}~\bibnamefont{Sato}}, \bibnamefont{and}
  \bibinfo{author}{\bibfnamefont{T.}~\bibnamefont{Takahashi}},
  \bibinfo{journal}{AIP Advances} \textbf{\bibinfo{volume}{1}},
  \bibinfo{pages}{022103} (\bibinfo{year}{2011}).

\bibitem[{\citenamefont{Holzwarth and Rabii}(1977)}]{Holzwarth1977}
\bibinfo{author}{\bibfnamefont{N.}~\bibnamefont{Holzwarth}} \bibnamefont{and}
  \bibinfo{author}{\bibfnamefont{S.}~\bibnamefont{Rabii}},
  \bibinfo{journal}{Mater. Sci. Eng.}
  \textbf{\bibinfo{volume}{31}}, \bibinfo{pages}{195} (\bibinfo{year}{1977}).

\bibitem[{\citenamefont{Kambe et~al.}(1979)\citenamefont{Kambe, Dresselhaus,
  Dresselhaus, Basu, McGhie, and Fischer}}]{Kambe1979}
\bibinfo{author}{\bibfnamefont{N.}~\bibnamefont{Kambe}},
  \bibinfo{author}{\bibfnamefont{M.}~\bibnamefont{Dresselhaus}},
  \bibinfo{author}{\bibfnamefont{G.}~\bibnamefont{Dresselhaus}},
  \bibinfo{author}{\bibfnamefont{S.}~\bibnamefont{Basu}},
  \bibinfo{author}{\bibfnamefont{A.}~\bibnamefont{McGhie}}, \bibnamefont{and}
  \bibinfo{author}{\bibfnamefont{J.}~\bibnamefont{Fischer}},
  \bibinfo{journal}{Mater. Sci. Eng.}
  \textbf{\bibinfo{volume}{40}}, \bibinfo{pages}{1} (\bibinfo{year}{1979}).

\end{thebibliography}

\end{document}